\documentclass[preprint,showpacs,preprintnumbers,amsmath,amssymb]{revtex4}
\usepackage{graphicx}
\usepackage{dcolumn}
\usepackage{bm}
\usepackage{epsfig}
\usepackage{xcolor}

\renewcommand{\r}{{\bf r}}

\newcommand{\be}{\begin{equation}}
\newcommand{\ee}{\end{equation}}
\newcommand{\beq}{\begin{eqnarray}}
\newcommand{\eeq}{\end{eqnarray}}
\begin{document}

\title{Pairing correlations  and effective mass}
\author{Satoshi Yoshida~$^{\rm a}$
and Hiroyuki Sagawa~$^{\rm b}$}
\address{$^{\rm a}$~Science Research Center, Hosei University \\
2-17-1 Fujimi, Chiyoda, Tokyo 102-8160,   Japan \\
$^{\rm b}$~Center for Mathematical Sciences, the University of Aizu \\
Aizu-Wakamatsu, Fukushima 965-8580,  Japan
}
\begin{abstract}
We study the effect of effective mass on pairing correlations in the ground
states of superfluid nuclei $^{124}$Sn and  $^{136}$Sn.
Various parameter sets of Skyrme interactions and relativistic Lagrangians are 
adopted
to study   pairing correlations across a wide range of effective mass.
It is shown that surface-type pairing interaction gives an  almost 
 constant pairing
gap  as a function of the effective mass, while  volume-type pairing
interaction shows rather strong dependence of the pairing gap  upon the
effective mass.
The local pair potentials of various effective interactions are also examined in relation to  the effective mass.
\end{abstract}

\maketitle

\section{Introduction}

Pairing correlations have been known as one of the most important 
correlations
in finite nuclei and also infinite nuclear matter ~\cite{BM58,RS-NS}.
Pairing correlations play an especially crucial role 
in providing realistic
descriptions  of the saturation properties and the excitation spectra of
open shell nuclei.
As an effective pairing interaction,  volume-type and surface-type
pairing interactions have often been  used in Hartree-Fock(HF)+BCS
  and HF+Bogoliubov (HFB)
calculations of medium-heavy and heavy nuclei ~\cite{Dob95,Bender}.
Recently, these pairing interactions have been 
   applied not only to stable nuclei
but also to unstable nuclei far from the stability line.
It has been known that the pairing correlations are much affected by the 
level
density around the Fermi energy, which is essentially determined by the
effective mass ~\cite{Mahaux85}.
The level density may also change from stable nuclei to unstable nuclei near
the drip lines where new shell closures   appear.

Many different parameter sets are available for mean field 
theories,   such as 
 Skyrme Hartree-Fock (SHF) and relativistic mean field (RMF) models.
These parameter sets show a large range of the 
 effective mass  $m^{*}/m =(0.4\sim1.0)$.
However, the effect of  effective mass on the pairing correlations has
not
yet  been studied   systematically  using  different mean field models.
In this article, we examine  how much the pairing correlations are influenced by 
the
effective mass of adopted interaction.
To this end, we take  two $\delta$-type pairing interactions,  
volume-type
and  surface-type,  and apply them to Sn isotopes in SHF+BCS and RMF+BCS   models.
Two isotopes,  $^{124}$Sn and $^{136}$Sn, 
 are chosen for  detailed
study since they are spherical and accurate  experimental data of 
  binding energies
 is  available.
 This paper is organized as follows. 
We give a brief summary of 
  the mean field models for the study of  pairing correlations 
in Section II.  Section III is devoted to discussion of  results of numerical 
calculations.   The summary is presented in Section IV.  

\section{mean field models for pairing correlations}

We study  the relation between the averaged pairing gap energy and the
effective mass in the SHF+BCS  and RMF+BCS models.
The Skyrme interaction $V_{Sky}$ is an effective zero-range force with 
density-
  and momentum-dependent terms \cite{Skyrme},
\begin{eqnarray}
V_{Sky}(\vec{r}_1,\vec{r}_2)&=&t_0 ( 1+x_0 P_{\sigma} ) \delta(\vec{r}_1
-\vec{r}_2)
+\frac{1}{2} t_1 (1+x_1 P_{\sigma}) \{ \vec{k}^{\prime 2}
\delta(\vec{r}_1-\vec{r}_2)+\delta(\vec{r}_1-\vec{r}_2) \vec{k}^2 \}
\nonumber \\
&&+t_2 (1+x_2 P_{\sigma}) \vec{k}^{\prime} \cdot \delta(\vec{r}_1
-\vec{r}_2) \vec{k} +\frac{1}{6} t_3 (1+x_3 P_{\sigma})
\rho^{\alpha}(\vec{r}) \delta(\vec{r}_1-\vec{r}_2) \nonumber \\
&&+ i W (\vec{\sigma}_1+ \vec{\sigma}_2) \cdot \vec{k}^{\prime}
\times \delta(\vec{r}_1-\vec{r}_2) \vec{k}, \label{Skym-int}
\end{eqnarray}
where $\vec{k}=(\vec{\nabla}_{1}-\vec{\nabla}_{2})/(2 i)$,  acting on the 
right, and 
$\vec{k}^{\prime}=-(\overleftarrow{\nabla}_{1}-\overleftarrow{\nabla}_{2})
/(2 i)$,  acting on the left,  are the relative-momentum 
  operators, $P_{\sigma}$  is
the spin exchange operator, $\vec{\sigma}$ is the Pauli spin matrix,  and
$\vec{r}=\frac{1}{2}(\vec{r}_1+\vec{r}_2)$.
The interaction (\ref{Skym-int}) simulates the G-matrix for nuclear 
Hartree
Fock calculations.
We use 13 different Skyrme parameter sets (SI, SIII, SIV, SVI, Skya, SkM,
SkM$^{*}$, SLy4, MSkA, SkI3, SkI4, SkX, SGII), taken from 
Refs.~\cite{Vautherin,Beiner,Kohler,Bohigas,Bartel,Reinhard,Chabanat,Sharma,Brown,Giai}.
The effective mass of SHF can be expressed analytically as
\be
   \frac{\hbar^2}{2m^*_q(\r)}= \frac{\hbar^2}{2m}
           +\frac{1}{8}\left [t_1(2+x_1)+t_2(2+x_2)\right ]\rho(\r)
           -\frac{1}{8}\left[t_1(1+2x_1)-t_2(1+2x_2)\right ]\rho_q(\r), 
\label{eq:eff-m1}
\ee
where $\rho_q(\r)$ is the neutron ($q=n)$ or proton density ($q=p)$,
and $\rho(\r)=\rho_n(\r)+\rho_p(\r)$.
In symmetric nuclear matter, Eq.~(\ref{eq:eff-m1}) gives
\begin{equation}
\frac{m^{*}_{\rm SHF}}{m}=\left[ 1+\frac{m \rho_{0}}{8 \hbar^2} \left\{ 3 
t_{1}+t_{2}
\left(5+4 x_{2} \right) \right\} \right]^{-1},
\end{equation}
where $\rho_{0}$ is the saturation density.
As RMF Lagrangians, we adopt the non-linear $\sigma$ model with  parameter 
sets NLSH, NL3,  and NLC \cite{Sharma2,Lalazissis,Serot}.
The effective mass of RMF can be expressed  as \cite{Ring96}
\be
 m^{*}_{\rm RMF}=m-\frac{1}{2}(V-S), 
\label{eq:eff-m}
\ee
where $V$ and $S$ are the vector and  scalar potentials.
For the RMF Lagrangian with $\sigma- $, $\omega- $, and $\rho-$meson fields, 
the
scalar potential is given by the  $\sigma-$potential, while the vector
potential depends upon the $\omega-,$ and $\rho-,$ and Coulomb potentials.
Note that the effective mass $m^{*}_{\rm RMF}$ in Eq.~(\ref{eq:eff-m}) is 
different
from  so-called ``Dirac mass'' given by $m^*_{\rm Dirac}=m+S$  \cite{Ring96}.

In this study, we adopt two types of pairing interactions,  
surface-  and  volume-type,  in the mean
field calculations.
The pairing interaction $V_{pair}$ is generally defined by the equation 
\begin{equation}
V_{pair}({\bf r})=-V_{0} \left[ 1- \beta \left (\frac{\rho({\bf r})}
{\rho_{0}}\right )^{\gamma} \right], 
\label{vpair}
\end{equation}
where $V_{0}, \beta$ and $\gamma$ are parameters to be determined 
   by the model.
The value of $\beta$ is fixed to be 0 and 1 
for  the volume-  and  surface-type pairing interactions, respectively.
We  also use the  values $\rho_{0}$= 0.16 fm$^{-3}$ and $\gamma$=1.
The pairing strength $V_{0}$ is  chosen to be 999~MeV$\cdot$fm$^{-3}$ for
   the surface-type pairing interaction and
323MeV~$\cdot$fm$^{-3}$ for   the volume-type~\cite{Bender}.
These pairing strength parameters were originally determined to reproduce
empirical pairing gaps of medium heavy nuclei with a Skyrme interaction 
SkI4.
We adopt these values as a reference in the following calculations with all
adopted Skyrme interactions and also RMF Lagrangians.
There is some freedom in the selection of values for  $\beta$ and $\gamma$.
It has been  pointed out that  the mixed-type pairing with  $\beta$=1/2
exhibits  good
properties to describe the pair density \cite{Jacek}.
A weak density dependence is also suggested by  the realistic 
nucleon-nucleon
interaction for the pairing channel \cite{Jerome07}.
Since it is not the main goal  of this article to study the effect of 
parameters $\beta$ and $\gamma$ on the pairing field, we adopt  the surface- 
and
volume-type pairing interactions which have been widely used in the
literature.

The pairing correlations are taken into account by BCS approximation 
in this study.  Since  
 the two-neutron separation energies  $S_{2n}$ of two nuclei 
 $^{124}$Sn and $^{136}$Sn
are more than 5~MeV,  BCS  approximation
is considered to be  a good model for 
  evaluating  the pairing correlations in the ground states 
 \cite{Grasso01,Chabanat}. 
The state-dependent gap strength $G_{ij}$ is defined by
\begin{equation}
G_{ij}=- \int d {\bf r}~V_{pair}({\bf r}) |\phi_{i}({\bf r})|^2 |\phi_{j}
({\bf r})|^2,
\end{equation}
where $\phi_{i}$ is a mean field single-particle wave function.
The pairing gap energy $\Delta_{i}$ is obtained by solving the following
BCS gap equation,
\begin{equation}
\Delta_{i}=\frac{1}{2} \sum_{j} f_{j} \frac{G_{ij}
\Delta_{j}}{\sqrt{(\varepsilon_{j}-\lambda)^2+\Delta_{j}^2}},
\end{equation}
where $\varepsilon_{j}$ and $\lambda$ are the  single particle 
and the Fermi energy, respectively,  and $f_{j}$ is a smooth cut-off factor 
for
the continuum states,  defined by
\begin{equation}
f_{j}=\frac{1}{1+e^{(\varepsilon_{j}-\lambda-\mu)/\Delta E}},
\end{equation}
where $\mu$ and $\Delta E$ are fixed to be 5.0 and 0.5, respectively.
The occupation probability $v_{i}^2$ is defined by
\begin{equation}
v_{i}^2=\frac{1}{2} \left[ 1- \frac{\varepsilon_{i}-
\lambda}{\sqrt{(\varepsilon_{i}-\lambda)^2+\Delta_{i}^2}} \right].
\end{equation}

We show the correlation  between the effective mass
and the averaged neutron pairing gap energy $\bar{\Delta}$ of $^{124}$Sn and
$^{136}$Sn in Figs.~\ref{eff-mass124} and \ref{eff-mass136}, respectively.
There are two definitions of the averaged  pairing gap energy 
$\bar{\Delta}$,
\begin{eqnarray}
\bar{\Delta}_{v^2}&=&\frac{\sum_{i \in n} v_{i}^{2} \Delta_{i}}{\sum_{i \in 
n}
v_{i}^{2}},
\label{v2} \\
\bar{\Delta}_{uv}&=&\frac{\sum_{i \in n} u_{i} v_{i} \Delta_{i}}{\sum_{i \in 
n}
u_{i} v_{i}},
\label{uv}
\end{eqnarray}
where $u_{i}=\sqrt{1-v_{i}^{2}}$.
In the next section, we discuss the averaged  pairing gap energies
$\bar{\Delta}_{v^{2}}$ and $\bar{\Delta}_{uv}$ for neutrons.

\section{Results of SHF+BCS and RMF+BCS calculations}

\begin{figure}[p]
\epsfig{file=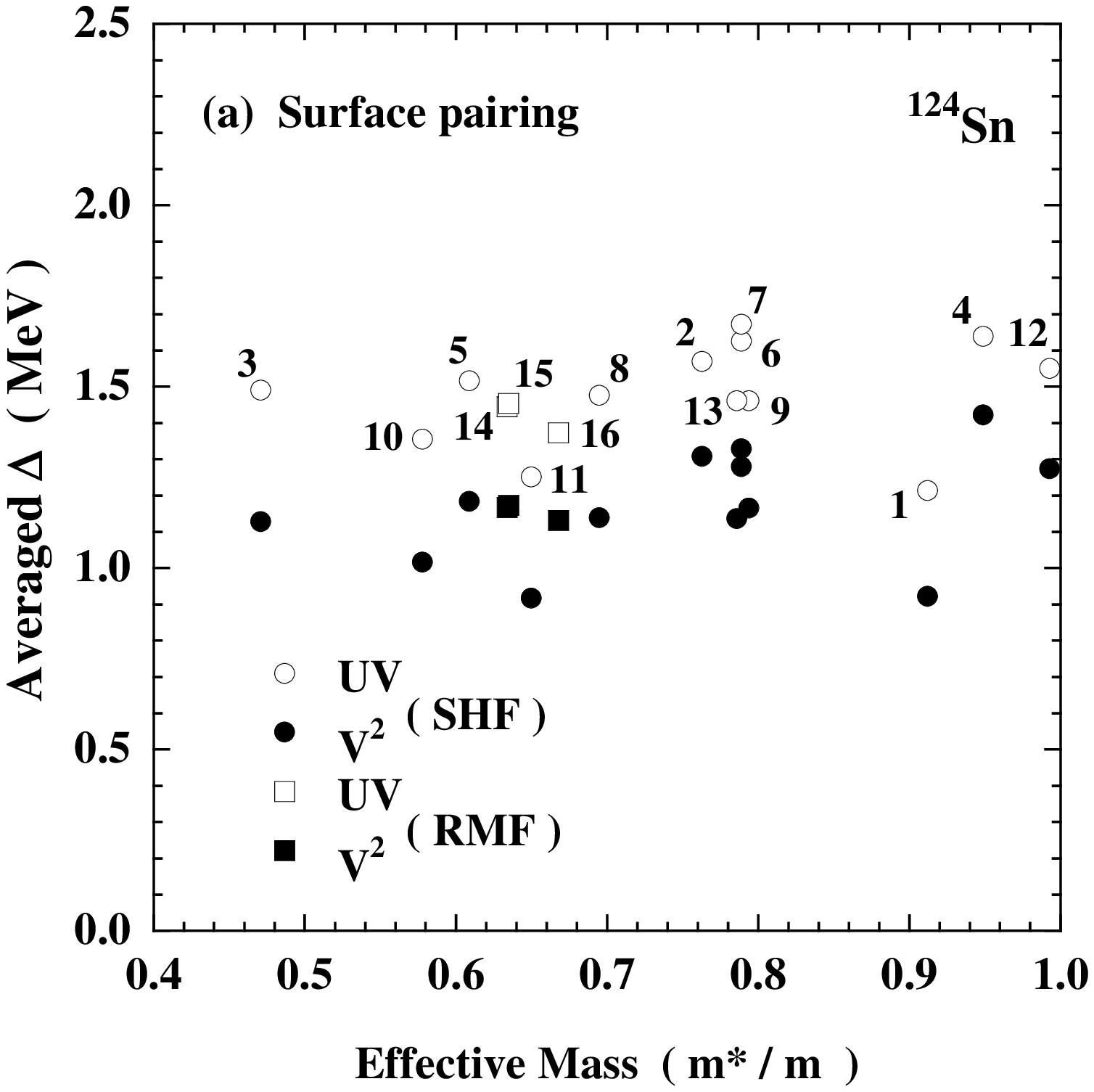,width=8cm}
\epsfig{file=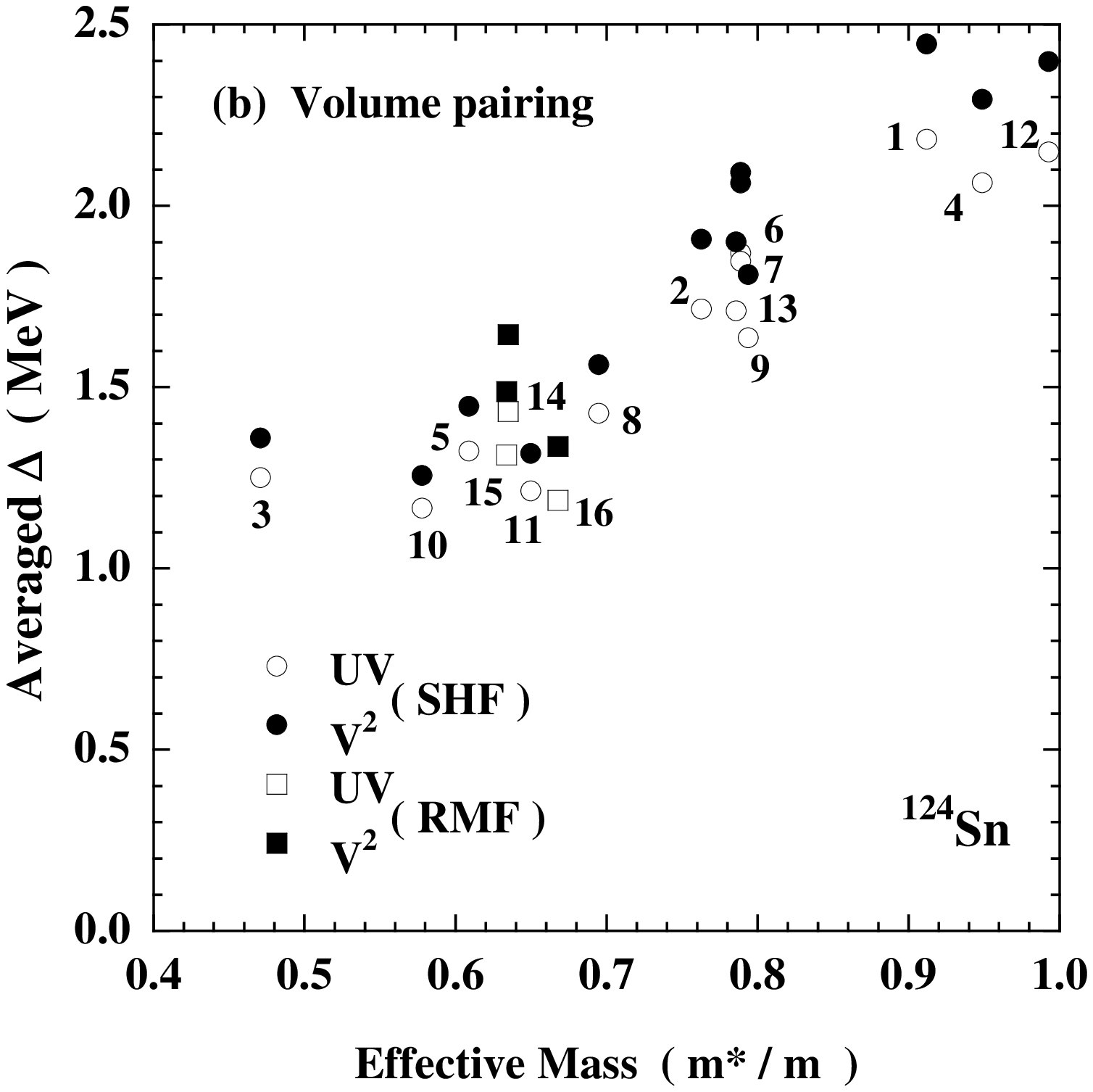,width=8cm}
\caption{The averaged neutron pairing gap energy $\bar{\Delta}$ and the
effective mass in $^{124}$Sn.  The left panel (a) is 
 for surface-type pairing interaction, while  the right panel (b) is 
for volume-type pairing interaction  in SHF+BCS and RMF+BCS
 models.  
Open circles (squares) and filled circles  (squares) 
 are obtained by averaging with
 $v^2$ and $uv$ factors in Eqs.\,(\ref{v2}) and (\ref{uv}), respectively,  in 
SHF(RMF)+BCS model.
The numbers denote the different parameter sets: 1 for SI, 2 for SIII, 3 for
SIV, 4 for SVI, 5 for Skya, 6 for SkM, 7 for SkM$^{*}$, 8 for SLy4, 9 for 
MSkA,
10 for SkI3, 11 for SkI4, 12 for SkX, 13 for SGII, 14  for
 NL3,  15 for NLC,  and 16 for NLSH.
}
\label{eff-mass124}
\end{figure}

\begin{figure}[p]
\includegraphics[width=3.2in,clip]{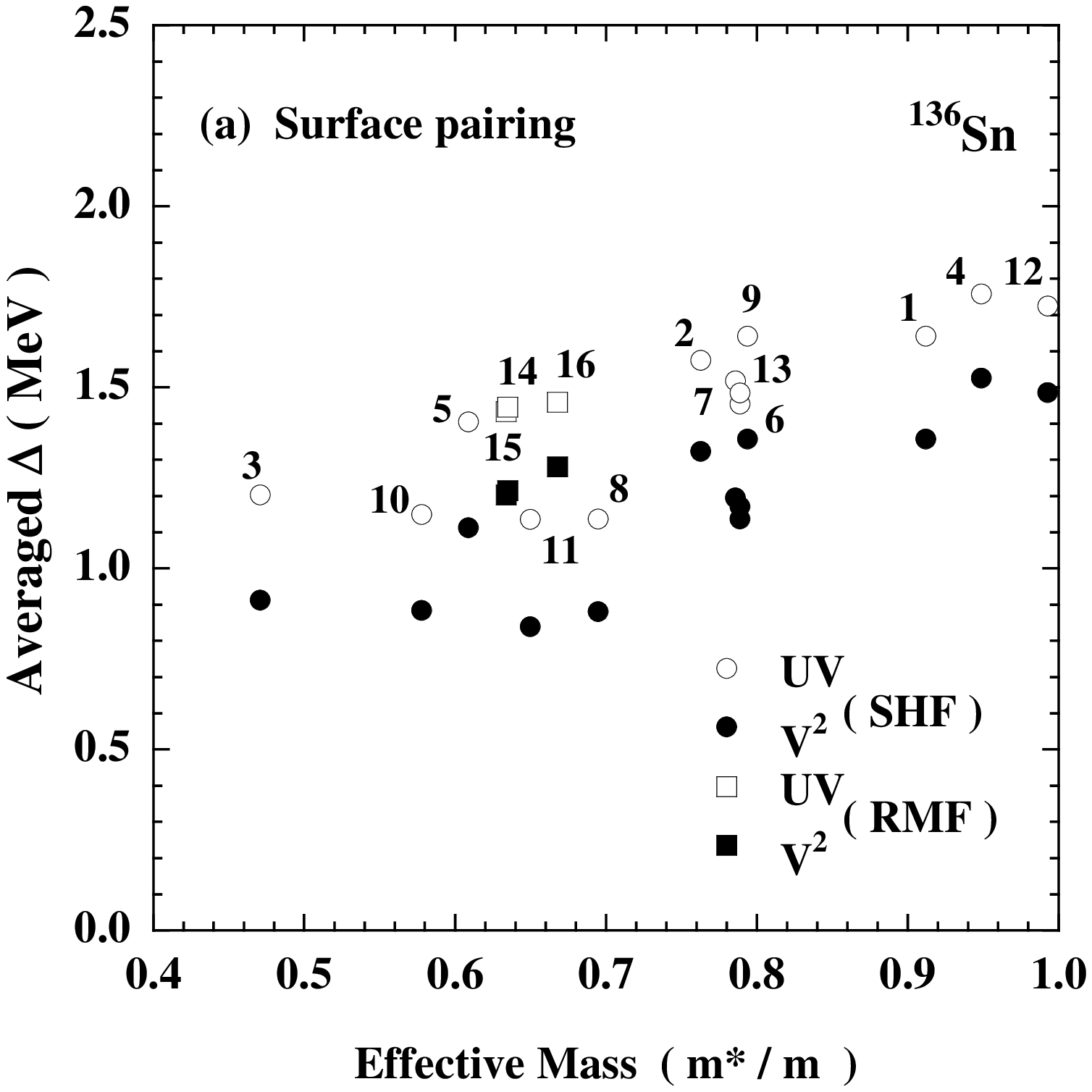}
\includegraphics[width=3.2in,clip]{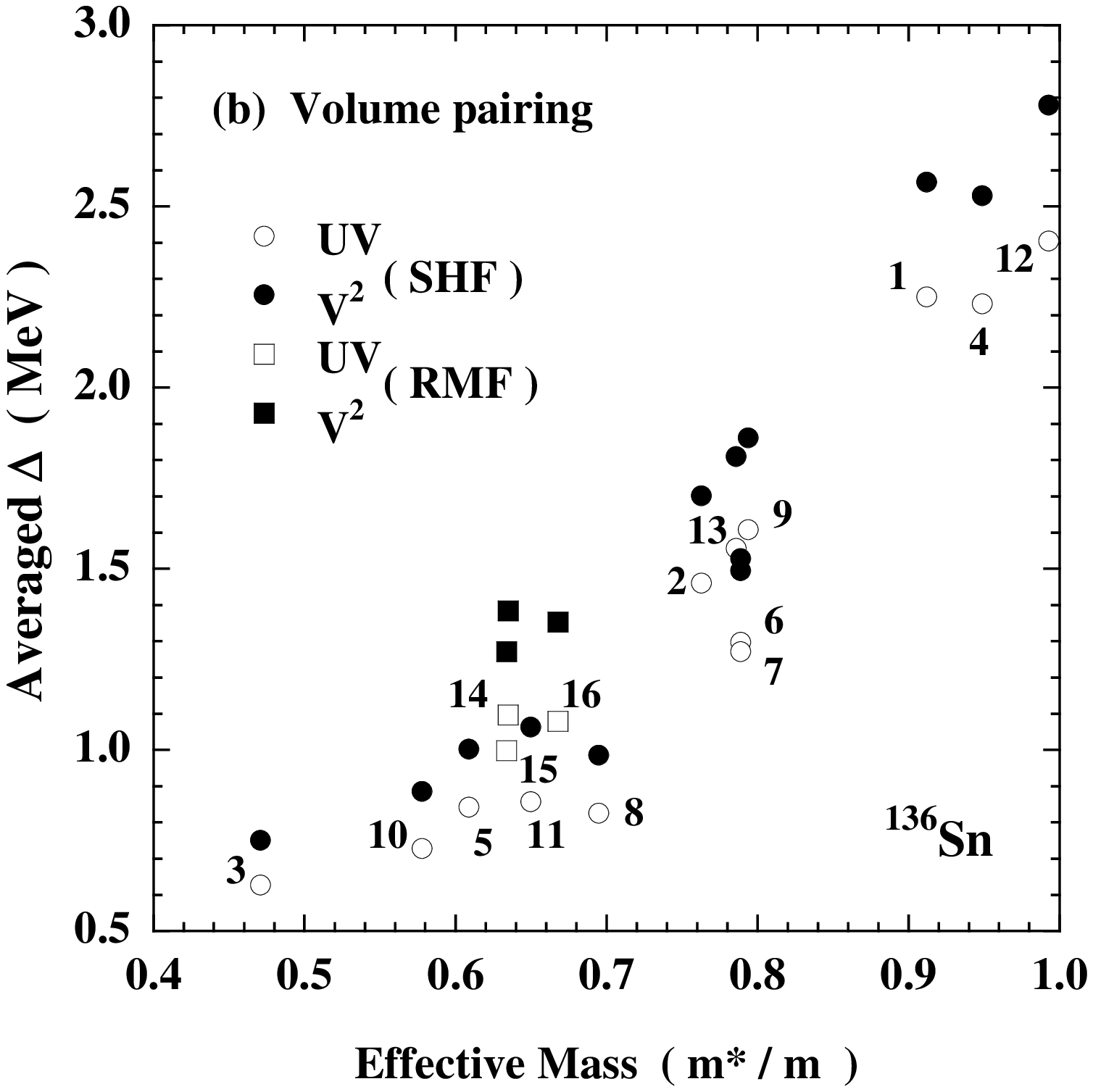}
\caption{The averaged neutron pairing gap energy $\bar{\Delta}$ and the
effective mass in $^{136}$Sn.   The left panel (a) is 
 for surface-type pairing  and  the right panel (b) is 
for volume-type pairing interaction   in SHF+BCS and RMF+BCS
 models.  
Open circles (squares) and filled circles  (squares) 
 are obtained by averaging with
 $v^2$ and $uv$ factors in Eqs.\,(\ref{v2}) and (\ref{uv}), respectively,  in 
SHF(RMF)+BCS model.
The numbers denote the different parameter sets: 1 for SI, 2 for SIII, 3 for
SIV, 4 for SVI, 5 for Skya, 6 for SkM, 7 for SkM$^{*}$, 8 for SLy4, 9 for 
MSkA,
10 for SkI3, 11 for SkI4, 12 for SkX, 13 for SGII, 14  for
 NL3,  15 for NLC, and 16 for NLSH.
}
\label{eff-mass136}
\end{figure}

As shown in Fig.~\ref{eff-mass124}(a), the averaged neutron pairing gap 
energies
$\bar{\Delta}_{v^{2}}$ and $\bar{\Delta}_{uv}$ with  the surface-type pairing
interaction are almost independent of  the effective mass with only minor 
variation.
Contrarily, the volume-type pairing shown in Fig.~\ref{eff-mass124}(b)
 has a strong
effective mass dependence. Specifically,  Fig.\,\ref{eff-mass124}(b) reveals  
 a correlation between the effective mass and the gap energy, i.e., 
 a larger effective mass yields  a larger  averaged neutron pairing
 gap energy.  
We  can see also that,  in the surface-type pairing interaction, the averaged
neutron pairing gap energies $\bar{\Delta}_{uv}$ are, on the average,
 300 keV larger than $\bar{\Delta}_{v^{2}}$. 
In contrast, the  pairing gap energies 
$\bar{\Delta}_{v^{2}}$
are 100 keV larger than  $\bar{\Delta}_{uv}$ in the volume-type pairing
interaction.
As shown in Fig.\,\ref{eff-mass136}, the results of  $^{136}$Sn have 
  almost the same
characteristic features  as those of $^{124}$Sn.
Quantitatively, the dependence of the averaged neutron pairing gap upon the effective mass in 
$^{136}$Sn  is somewhat stronger than that of 
$^{124}$Sn.
 No 
substantial difference is observed between SHF and RMF in Figs.
\ref{eff-mass124} and \ref{eff-mass136} regarding 
the correlation between
$\bar{\Delta}$ and the effective mass.

\begin{figure}[p]
\includegraphics[width=5.5in,clip]{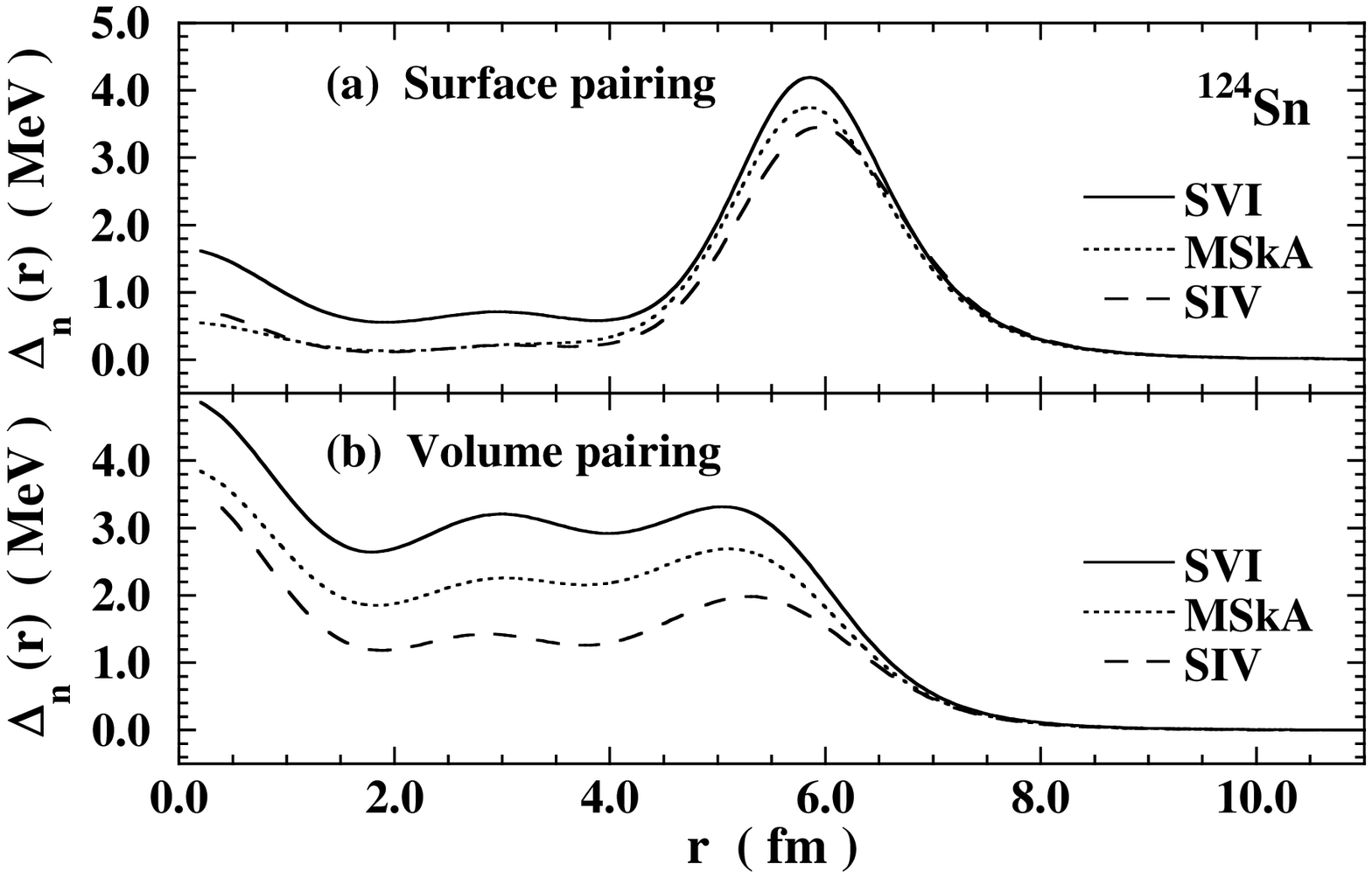}
\caption{Local pair potentials of $^{124}$Sn in SHF+BCS model.
(a) with surface-type and (b) with volume-type pairing interactions.
Solid, dotted,  and dashed lines are results for SVI, MSkA and SIV
parameter sets, respectively.  See the text for details.  
}
\label{pair-potential124}
\end{figure}

\begin{figure}[p]
\includegraphics[width=5.5in,clip]{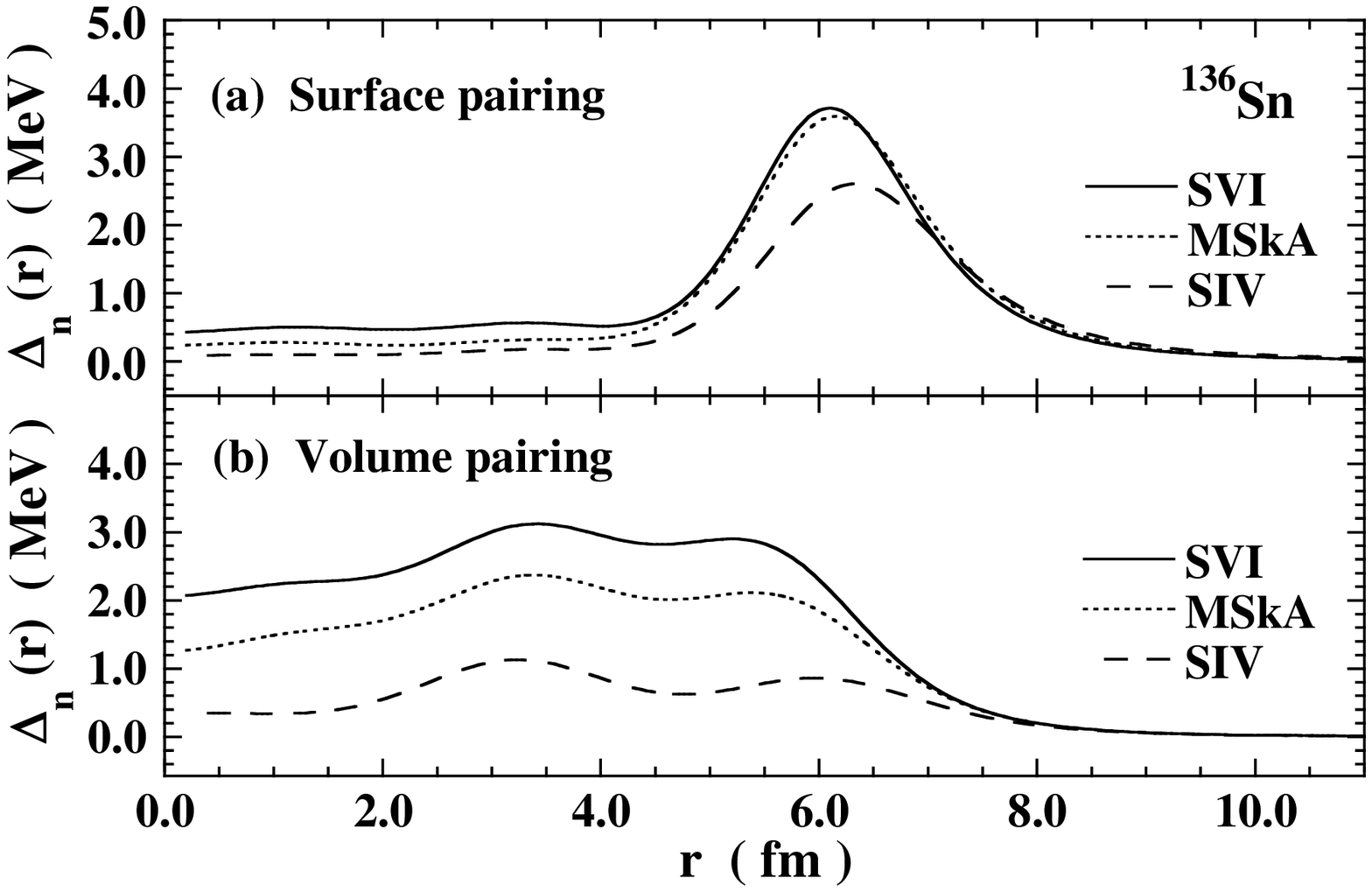}
\caption{Local pair potentials of $^{136}$Sn in SHF+BCS model.
(a) with volume-type and (b) with surface-type pairing interactions.
Solid, dotted,  and dashed lines are results for SVI, MSkA,  and SIV
parameter sets, respectively. See the text for details.
}
\label{pair-potential136}
\end{figure}

Figures \ref{pair-potential124} and \ref{pair-potential136} show the local
pairing potentials of neutrons $\Delta_{n}(r)$ for $^{124}$Sn and 
$^{136}$Sn,
respectively, in the coordinate space.
The local pair potential of neutrons is given  by
\begin{equation}
\Delta_{n}(r)=-V_{pair}(r) \sum_{i \in \Omega_{n},i>0} u_{i} v_{i} | 
\phi_{i}
(r) |^2.
\label{local-pot}
\end{equation}
The upper panels show the pair potentials of  surface-type pairing, while
the lower panels show those of volume-type pairing with the Skyrme
interactions SVI ($m^{*}/m=0.949$), MSkA ($m^{*}/m=0.794$),  and SIV ($m^{*}/m
=0.471$).
These interactions are denoted by the numbers 4 (SVI), 9 (MSkA),  and 3 (SIV),
respectively,  in Figs.\,\ref{eff-mass124} and \ref{eff-mass136}.
The local pair potential $\Delta_{n}(r)$ gives the pairing gap energy
$\Delta_{i}$ for a single particle state $i$ as
\begin{equation}
\Delta_{i}=\int d {\bf r} \phi^{\dagger}_{i}({\bf r}) \Delta_{n} ({\bf r})
\phi_{i}({\bf r}).  
\end{equation}
As shown in Figs.\,\ref{pair-potential124} and \ref{pair-potential136},
the graph of $\Delta_n(r)$ with  the volume-type interaction has
a plateau inside the nuclear surface similar to the neutron density 
 $\rho_n(r)$, whereas the graph of $\Delta_n(r)$ with the 
surface-type pairing interaction has a bump at  the nuclear surface.
In general, the value of $u_{i} v_{i}$ reaches a  maximum for the single
particle states close to the Fermi level, whereas the value of $v_{i}^2$ is
unity
for the deeply bounded single particle states below the Fermi level and
decreases near the Fermi level,  eventually, vanishing  in the continuum.
The averaged neutron pairing gap energies $\bar{\Delta}_{v^2}
(\bar{\Delta}_{uv}$) are  obtained by averaging $\Delta_{i}$ with the factor
$v_{i}^2$ ($u_{i} v_{i}$) over all single-particle states in Eq.\,(\ref{v2})
(Eq.\,(\ref{uv})).
In the case of the surface-type pairing interaction, the pairing gap 
$\Delta_{i}$
is larger for the single-particle state near the Fermi surface since the
overlap of the surface peaked $\Delta_{n}(r)$ and the wave function $\phi_i$
is larger.
Then, the averaged neutron pairing gap energy $\bar{\Delta}_{uv}$ is larger
than $\bar{\Delta}_{v^2}$  because the product  $u_{i} v_{i}$ 
reaches its 
maximum for the single-particle states near the Fermi level.
In contrast, 
in the case of  volume-pairing interaction, the well-bound
single-particle states below the Fermi level have large  pairing gaps 
$\Delta_{i}$ because  the interior part of
$\Delta_{n}(r)$ has a large overlap with the well-bound wave functions.   
Then, for   the volume-type pairing interaction, $\bar{\Delta}_{v^2}$ becomes
larger than $\bar{\Delta}_{uv}$ because the occupation probability $v^2$ is
larger for the well-bound states.

The neutron densities of $^{124}$Sn and $^{136}$Sn are  a half of 
the respective  averaged central densities  at  $r=5.7$ and  
$5.9$ fm, respectively, 
irrespective of the Skyrme interactions SVI, MSkA and SIV.  We refer  to 
these radii as the surface positions of two nuclei.  
Let us now examine  the magnitude of the pair  potential  at this surface 
 position.  
In the case of surface-type pairing interaction shown 
in Fig.\,\ref{pair-potential124},  the difference
of the local pair potentials at $r=5.7$ fm  is only 19\% between SVI and SIV 
for $^{124}$Sn, 
while  it is 33\% in the volume-type pairing interaction.
The local pair potential  of MSkA stays  between those of SVI and SIV.
It turns out that the difference of $\bar{\Delta}_{uv}$ between SIV and SVI 
is
only 9\% in the surface-type pairing interaction, while  it  is 39\% in 
the
volume pairing interaction,  as shown in Fig.\,\ref{eff-mass124}.
Figure \ref{pair-potential136} for $^{136}$Sn shows that, in the case of
surface-type pairing interaction, the difference of the local pair 
potentials
at $r=5.9$ fm   is 36\% between SVI and SIV,  while  it  is 65\% in the
volume-type pairing interaction.
The corresponding 
  difference appears also in Fig.\,\ref{eff-mass136}:  the difference 
of
$\bar{\Delta}_{uv}$ between SIV and SVI is 32\% in the  surface-type pairing
interaction while  it  is 72\% in the  volume-type pairing interaction.
Thus, the variation of the averaged neutron pairing gap energy for different
effective masses  is much smaller for  the  surface-type pairing 
interaction
than for  the volume-type pairing interaction,  as seen in
 Figs.\,\ref{eff-mass124}, and \ref{eff-mass136}.
This phenomenon can be understood by the following discussion.

The single-particle states near the Fermi level claim  the dominant 
contribution 
to the averaged pairing gap energy  in the case of surface-type pairing, 
while
the deeply-bound single-particle states  also have substantial contributions 
in
the case of volume-type pairing. Since the single-particle levels in 
$^{124}$Sn
near the Fermi level are not much affected by  change of effective mass, 
the
averaged pairing gap energies in Fig. \ref{eff-mass124}(a) show only small
variation.
In contrast,
the single-particle energies of the well-bound states are 
very
much affected by the effective mass.
 Consequently,  the average pairing gaps in  Fig.\,\ref{eff-mass124}(b) 
 show a clear effective mass dependence due to
 the  well-bound states. 
These arguments apply  qualitatively in the case of $^{136}$Sn,  as shown in
Fig. \ref{eff-mass136}.
However, one can see larger 
  variation  in the  unstable nucleus $^{136}$Sn,  whose
single-particle states are more sensitive to  change of effective mass 
than the 
stable nucleus  $^{124}$Sn.

\begin{figure}[p]
\includegraphics[width=5.5in,clip]{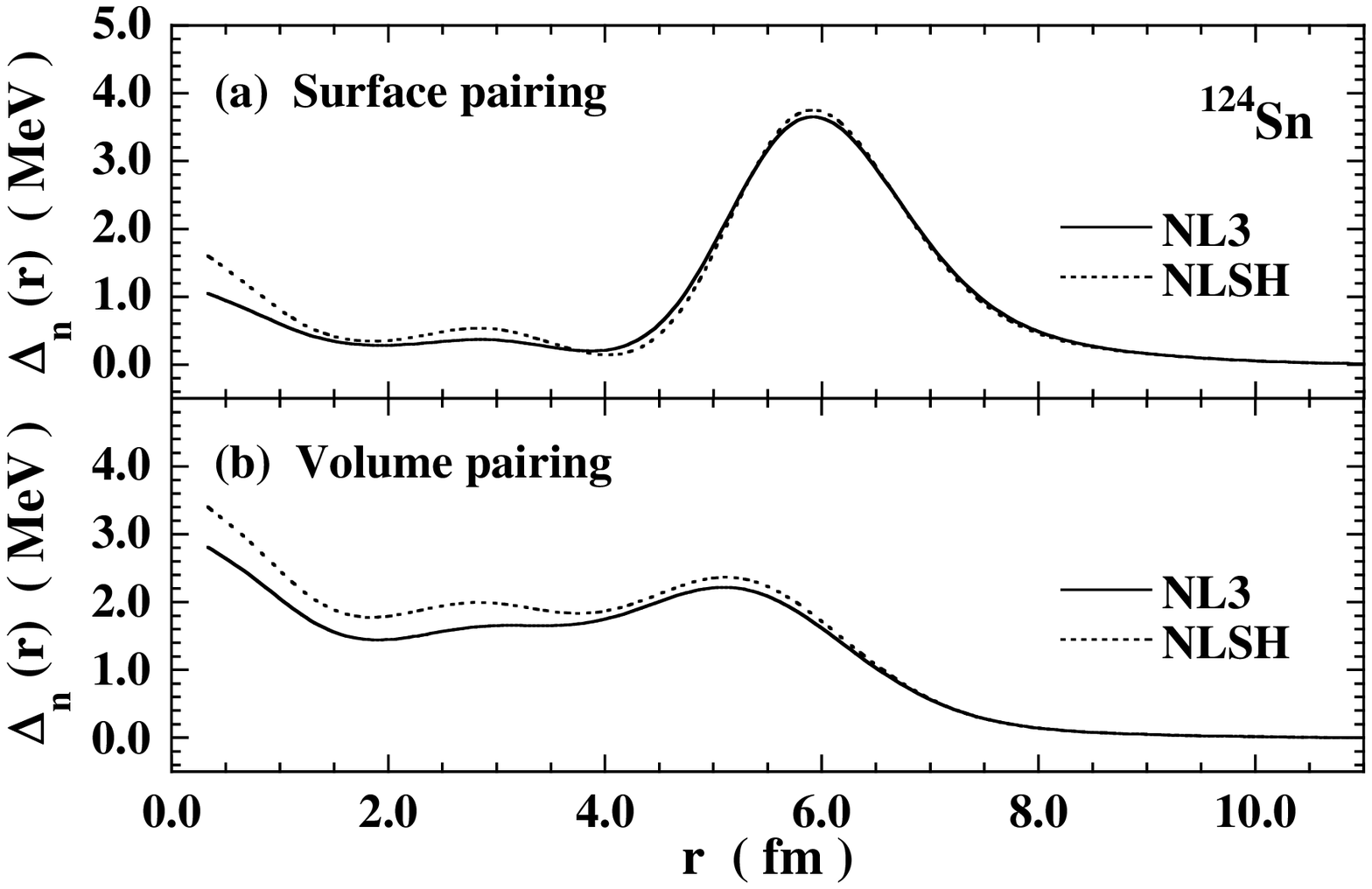}
\caption{Local pair potentials of $^{124}$Sn in RMF+BCS model.
(a) with surface-type and (b) with volume-type pairing interactions.
Solid and dotted lines are the results for NL3 and NLSH
parameter sets, respectively. See the text for details.
}
\label{rmf-pair-potential124}
\end{figure}

\begin{figure}[p]
\includegraphics[width=5.5in,clip]{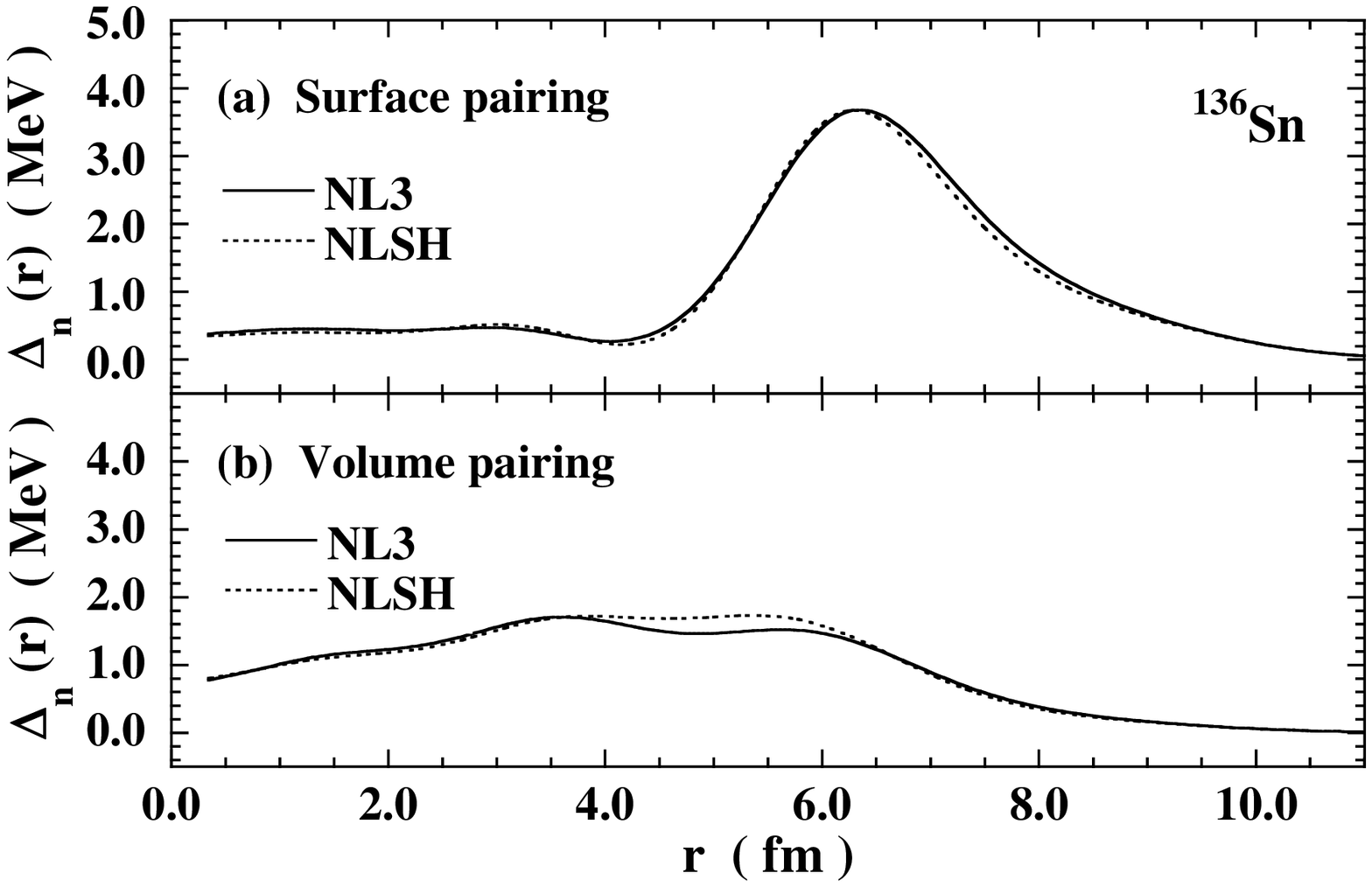}
\caption{Local pair potentials of $^{136}$Sn  in RMF+BCS model.
(a) with volume-type and (b) with surface-type pairing interactions.
Solid and dotted lines are the results for NL3 and NLSH
parameter sets, respectively. See the text for details.
}
\label{rmf-pair-potential136}
\end{figure}

The local pair potentials of RMF+BCS model are shown in 
Figs.\,\ref{rmf-pair-potential124} and \ref{rmf-pair-potential136}
  for $^{124}$Sn
and $^{136}$Sn, respectively,  with the RMF Lagrangians NL3 ($m^*/m$=0.634) 
and
NLSH ($m^*/m$= 0.635).
General features of RMF+BCS results are similar to those of SHF+BCS 
in Figs.\,\ref{pair-potential124} and \ref{pair-potential136}:
 the pair potential peaks  at the surface 
  in the case of  surface-type pairing
interaction, while the potential  plateaus within  the nuclear surface for 
the  volume-type pairing.
Because of the thick neutron skin, the surface peak is more extended in
$^{136}$Sn,  as  seen in the upper panel of Fig. 
\ref{rmf-pair-potential136}.
Comparing the two RMF Lagrangians NL3 and NLSH, the results appear very close
 to each other for the surface-type pairing,  while NL3 has a slightly larger pair
potential than NLSH for the volume-type pairing.
The averaged pairing gap potentials show a smooth trend as a function of
the effective mass in Figs.\,\ref{eff-mass124} and \ref{eff-mass136} where
both RMF and SHF results are plotted.  
This is due to  the similarities of the 
local pair potentials between the two models.

\section{Summary}
We studied  pairing correlations in the ground states of typical 
superfluid
nuclei $^{124}$Sn and $^{136}$Sn in relation to  the    effective 
mass in  SHF+BCS and RMF+BCS models. 
We adopted various parameter sets 
   of Skyrme interactions and relativistic Lagrangians together 
  with the surface- and volume-type 
pairing interactions to study  pairing correlations across
   a wide range of effective mass.  We showed 
 that the surface-type pairing interaction gives  an almost constant
pairing gap, independent of the effective mass, while the volume-type 
pairing
interaction shows  rather strong dependence of the pairing correlations
  upon the
effective mass.
The local pair potential in the case of surface-type pairing interaction
has a peak at the  surface of neutron density, while the potential 
 in the case of   
volume-type pairing interaction has a plateau inside the nuclear surface.
These features 
 of the local pair potentials cause 
 a  difference between the surface- and volume-type pairing interactions 
 regarding  effective mass dependency of the averaged neutron pairing gap energy.  
We  also 
  pointed out  that the effect of effective mass on the averaged pairing 
gap
energies is essentially the same between  RMF and SHF models since 
   the local pair potentials have similar characteristic features
 in the two models.


\end{document}